# A deep learning-based framework for segmenting invisible clinical target volumes with estimated uncertainties for post-operative prostate cancer radiotherapy

Anjali Balagopal[1], Dan Nguyen[1], Howard Morgan, Yaochung Weng, Michael Dohopolski, Mu-Han Lin, Azar Sadeghnejad Barkousaraie, Yesenia Gonzalez, Aurelie Garant, Neil Desai, Raquibul Hannan, Steve Jiang[*]

**Medical Artificial Intelligence and Automation Laboratory and Department of Radiation Oncology, University of Texas Southwestern Medical Center, Dallas, Texas, USA**
Email: Steve.Jiang@utsouthwestern.edu

**Abstract.** In post-operative radiotherapy for prostate cancer, the cancerous prostate gland has been surgically removed, so the clinical target volume (CTV) to be irradiated encompasses the microscopic spread of tumor cells, which cannot be visualized in typical clinical images such as computed tomography or magnetic resonance imaging. In current clinical practice, physicians segment CTVs manually based on their relationship with nearby organs and other clinical information, per clinical guidelines. Automating post-operative prostate CTV segmentation with traditional image segmentation methods has been a major challenge. Here, we propose a deep learning model to overcome this problem by segmenting nearby organs first, then using their relationship with the CTV to assist CTV segmentation. The model proposed is trained using labels clinically approved and used for patient treatment, which are subject to relatively large inter-physician variations due to the absence of a visual ground truth. The model achieves an average Dice similarity coefficient (DSC) of 0.87 on a holdout dataset of 50 patients, much better than established methods, such as atlas-based methods (DSC<0.7). The uncertainties associated with automatically segmented CTV contours are also estimated to help physicians inspect and revise the contours, especially in areas with large inter-physician variations. We also use a 4-point grading system to show that the clinical quality of the automatically segmented CTV contours is equal to that of approved clinical contours manually drawn by physicians.

Radiotherapy is typically considered for patients with localized prostate cancer who have already undergone prostatectomy. For patients with adverse pathologic findings, adding radiotherapy after radical prostatectomy reduces the risk of biochemical recurrence (as measured by the level of prostate-specific antigen or PSA, a commonly used surrogate for prostate cancer recurrence), local recurrence, and clinical progression of cancer over surgery alone. Salvage radiotherapy is recommended if patients show an increase in PSA or local recurrence after prostatectomy [1, 2].

Optimal radiation treatment entails uniform full-dose coverage of the radiation target with a sharp dose fall-off around it. This necessitates precise segmentation of both the radiation target and the nearby organs that are at risk for radiation damage (Organs-at-risk, OARs). In a typical radiotherapy setting, OARs are segmented together with the gross tumor volume (GTV), which is the tumor that is visible in images. Using their knowledge of the disease, physicians then expand the GTV to create the clinical target volume (CTV), which also includes the microscopic extensions not visible in images. In the case of post-operative radiotherapy for prostate cancer, however, the prostate gland has been surgically removed, so the CTV is only a virtual volume encompassing areas that may contain microscopic tumor cells, not an expansion of a macroscopic or visible tumor volume. Consequently, segmenting the CTV in this case, whether manually or automatically, is much more challenging than segmenting typical organs or CTVs expanded from GTVs.

---
[1] *Co-first authors*

Four published consensus guidelines have variously defined the post-operative CTV in prostate cancer radiotherapy (European Organization for Research and Treatment of Cancer [EORTC], Faculty of Radiation Oncology Genito-Urinary Group [FROGG], Princess Margaret Hospital [PMH], and Radiation Therapy Oncology Group [RTOG]) [3-6]; however, there is no universally accepted method of segmenting the CTV. The CTV's borders in each of these guidelines are based on anatomical landmarks, and there are important differences between guidelines. It is also recognized that physicians must consider not only published guidelines, but individual patient characteristics, variability in anatomy, and co-morbidities, when segmenting the CTV. There is a lot of room for physicians to exercise their best clinical judgment for individual patients, based on their training backgrounds, experiences, and personal preferences, which can lead to large variations among physicians [7-11].

Accurately segmenting the CTV is crucial for irradiating the microscopic spread of tumor cells sufficiently while mitigating the side effects of radiation therapy to surrounding OARs. Traditional methods for automatic image segmentation, while suitable for OARs and GTVs, cannot be used for post-operative prostate CTV segmentation, because the CTV in this case is a virtual volume whose boundaries are not defined by the grayscale variation in the images, and the CTV is not an expansion of a GTV [12]. Among traditional image segmentation methods, atlas-based methods are considered the best suited for solving this problem. However, these methods have not produced satisfactory results, as they achieved a Dice similarity coefficient (DSC) of less than 0.7 for post-operative CTV segmentation [13].

In recent years, deep learning (DL) methods have produced better results than conventional methods in medical image segmentation. Multiple DL studies have focused on the automatic segmentation of OARs and GTVs [14-20]. Elguindi et al. [21] have effectively used a DL model to automatically segment pre-operative prostate CTVs on magnetic resonance imaging (MRI). Men et al. have used DL models to segment CTVs for pre-operative rectal cancer [22] and nasopharyngeal cancer [23]. These studies deal with pre-operative CTV segmentation, where the GTV is visible. Recently, Song et al. [25] applied off-shelf DL models [26] to segment CTVs for post-operative rectal cancer in a way similar to DL-based OAR segmentation. Although the results are acceptable (DSC=0.88) for rectal cancer, their approach may not be applicable to segmenting the CTVs of other tumor sites. With post-operative rectal cancer CTVs, the volumes are typically very large, and it is easy to achieve a high DSC value. In addition, the volumes are well defined by surrounding anatomical landmarks, including the external anal sphincter, inferiorly; the pelvic floor musculature, sacrum, and lumbar vertebrae, posteriorly; the pelvic floor musculature and sacroiliac joints, laterally; approximately 5 mm into the bladder or 7 mm around vessels, anteriorly; and up to the aortic bifurcation, superiorly. Also, this study utilized the contours of only one radiation oncologist which would not capture the inter-physician variability of post-operative CTVs seen in everyday practice, potentially inflating the DSC score further. For other tumor sites, such as prostate cancer, the volumes of post-operative CTVs can be much smaller, and the boundaries are far less defined by anatomical landmarks. Therefore, a straightforward application of DL-based OAR segmentation models is unlikely to work, as special architectures and techniques for the most efficient feature learning would be required. Also, post-operative CTVs vary greatly between physicians, so this variability must also be addressed by segmentation models.

In this work, we sought to address the clinical need for fast, accurate, and consistent CTV segmentation for post-operative prostate cancer radiotherapy by proposing a DL-based workflow that automatically segments CTVs based on post-operative prostate cancer CT images. The CTV contour for this application is based on complicated anatomical features and has no boundaries. The size of the CTV varies significantly but is typically much smaller than post-operative rectal cancer CTVs. For automatic CTV segmentation, meaningful information has to be extracted from physician-labeled contours, which are based on the anatomical borders of nearby OARs, human perception and experience, and other clinical variables. Because the anatomical borders of OARs play such an important role in accurately segmenting the CTV, we leverage automatically segmented OAR contours to automatically segment the CTV (Anatomy Guided, AG). As anatomical location information also plays an important role in accurate CTV segmentation, we use a multi-task network (MTN) that predicts, as an auxiliary task, a distance map showing the minimum distance of each voxel to the CTV boundary. To account for the different shapes, sizes, and intensities between the bladder, rectum, penile bulb and femoral heads, we selected different existing network architectures to best segment each OAR.

Contours produced by DL models inevitably have large uncertainties in particular areas because of the absence of clear boundaries in the images, variations in image quality, low soft tissue contrast in CT images, the model's own ignorance of the data, and large inter-physician variations, especially for post-operative prostate CTVs. Quantifying model uncertainties could yield important information regarding errors in prediction and failures on unseen data when the model is used in clinical settings. Such information could be valuable for physicians as they review, revise, and approve the CTV contours that the model produces.

The key to estimating model uncertainty is the posterior distribution $p(W|X,Y)$, also referred to as Bayesian inference. There have been various research efforts on approximating Bayesian inference in DL. A recent insight from Gal and Ghahramani [27], termed Monte Carlo dropout (MCDO), is particularly promising for use in medical settings because it is easy to use and inherently scalable. Using MCDO, one can efficiently approximate Bayesian inference to estimate the posterior distribution. Using MCDO to estimate uncertainty in DL was originally proposed for classification tasks [27] and later applied to semantic segmentation in computer vision [28], where it provides a pixel-wise estimate of model uncertainty. Here, we use MCDO to estimate the uncertainties associated with predicted CTV contours.

In this study, we propose an end-to-end method for automated 3D post-operative prostate CTV segmentation with uncertainty estimation. We quantitatively evaluate the predicted contours by comparing them with *clinical contours*, defined as contours used in clinical practice for patient treatment. These clinical contours are usually drawn by residents, corrected by supervising attending physicians, and reviewed by all attending physicians in the genitourinary radiation oncology service within our institution. We evaluated the performance of our DL model versus residents in terms of generating CTV contours by using the clinical contours as references. Attending physicians performed a separate evaluation by using a four-point grading system to estimate the clinical acceptability of the DL contours.

The proposed DL-based framework is fully automated (**Figure 1b**). It localizes structures in each patient's CT image and crops it into a volume of interest (VOI) for each structure. Then, it segments each OAR by using the corresponding VOI as input to a dedicated DL model. Last, it uses the segmented bladder and rectum together with the CTV VOI as inputs for CTV segmentation.

## Automation workflow

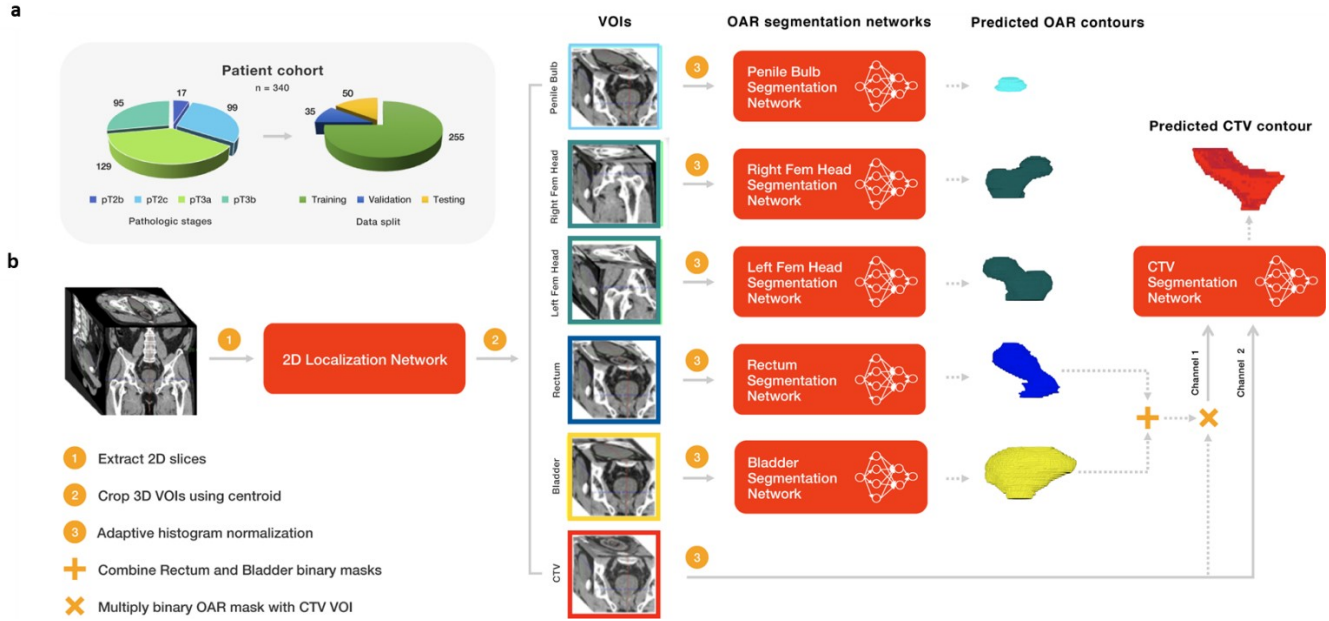

**Figure 1.**
*Diagram of the proposed workflow. **a**, Pie charts break down the patient cohort (340 patients) into pathologic stages of the tumor (17 pT2b, 99 pT2c, 129 pT3a, and 95 pT3b) and also into training (255), validation (35), and testing (50) sets. For each patient, the CT volume contains 60-360 slices and a voxel size of 1.17 × 1.17 × 3 $mm^3$. **b**, Fully automated workflow: CTV and OAR volumes are localized and cropped from the original CT images through a 2D localization network; OARs are segmented individually by separate 3D segmentation networks; and CTV is segmented by a dedicated 3D segmentation network that takes the localized CTV volume and segmented bladder and rectum as inputs.*

## Evaluation of the proposed workflow

**Quantitative Evaluation.** We quantitatively evaluated the contours predicted by the model against clinical contours by using two metrics: (1) volumetric DSC to evaluate the overlapping volume between the predicted and the clinical contours and (2) average surface distance (ASD) to evaluate the distance between the surfaces of the two contours. The test performance of the networks is summarized in **Figure 2**. Model-predicted CTV contours show high agreement with the clinical contours in terms of both DSC (87%) and ASD (1.6 mm). The model performed as well when predicting OAR contours as the state-of-the art DL models for the same organs [14,18,29]. We also implemented DeepLabv3+ network used for post-operative rectal CTV contouring [30, 25] and tested it for our problem. We found that the result (DSC score 79.4%) is much worse than that of the proposed AGMTN.

**Ablation Study.** We performed ablation studies to evaluate the impact of the multi-task network and anatomy guidance. We compared the DSC values of CTVs by using paired two-sided *t*-tests for the following network architectures: (1) Anatomy-Guided Multi-Task Network (AG-MTN), (2) Multi-Task Network (MTN), (3) Anatomy-Guided 3D UNet (AG-UNet), and (4) a plain 3D UNet [40]. The details of these models are given in the Methods section. The proposed AG-MTN outperformed the other networks, and the difference was statistically significant (**Figure 2**). The models that use distance prediction as an auxiliary task (Multi-Task) outperformed the models without distance prediction by about 3% (MTN versus UNet, and AG-MTN versus AG-UNet). The models that use bladder and rectum contours as model inputs (anatomy-guidance) outperformed the models without anatomy guidance by about 5% (AG-UNet versus UNet, and AG-MTN versus MTN). The AG-MTN model, which uses both techniques (multi-task and anatomy-guidance) outperformed the UNet model, which uses neither, by 8% on average.

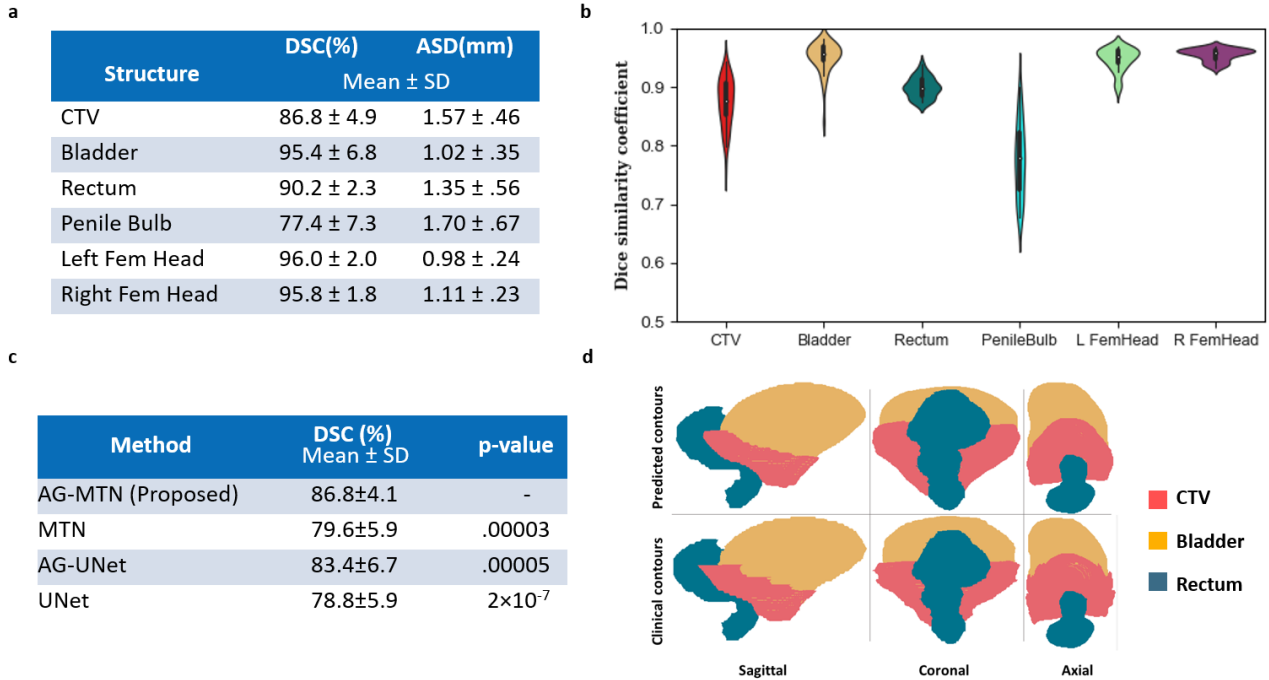

**Figure 2.**

*Quantitative evaluation of the predicted CTV and OAR contours against the clinical contours. **a**, Mean values and standard deviations of Dice similarity coefficient (DSC) and average surface distance (ASD); **b**, Violin plots of DSC values; **c**, Comparison of CTV DSC values from the ablation study for the proposed Anatomy-Guided Multi-Task Network (AG-MTN) against the Multi-Task Network (MTN), Anatomy-Guided 3D UNet (AG-UNet), and plain 3D UNet; **d**, Visualization of the predicted and clinical CTV, bladder, and rectum contours for an example test patient.*

**Uncertainty Evaluation.** MCDO operates by leaving Dropout on during the testing phase, where a portion of the network is randomly dropped for each prediction. Fifty MCDO samples were generated for CTV and OAR segmentation results for each patient, and we used these samples to calculate the mean and variance of the probabilities of each voxel belonging to a structure. The predicted mean contours and the associated 95% confidence bounds for CTVs are presented in **Figure 3**. The predicted contours for OARs can also be presented in the same way. When reviewing and approving the model-generated contours, physicians can focus on the areas of the contours with wide 95% confidence bounds.

The large uncertainties in the predicted CTV contours are mainly related to three clinical scenarios. First, the institutional guidelines provide a lot of leeway for segmenting the inferior CT slices, which results in large variations in physicians' practice (**Figure 3d**). Second, the large uncertainties in the middle CT slices come from the large variations in physicians' experience and preference in handling bladder toxicity (**Figure 3e, g, h**). Third, the variability in physicians' choices to include or exclude seminal vesicle remnants leads to large CTV contour uncertainties in the superior CT slices (**Figure 3i-l**).

When the estimated uncertainty is small, as in **Figure 3a-c and f**, the model-predicted contours match well with the clinical contours. In areas where the uncertainty is large, the predicted contours may (**Figure 3e, i**) or may not (**Figure 3h, l**) agree well with the clinical contours, so the 95% confidence bounds could be a useful tool that helps physicians to inspect and revise contours in an informed, efficient way when using the developed DL model in clinical practice.

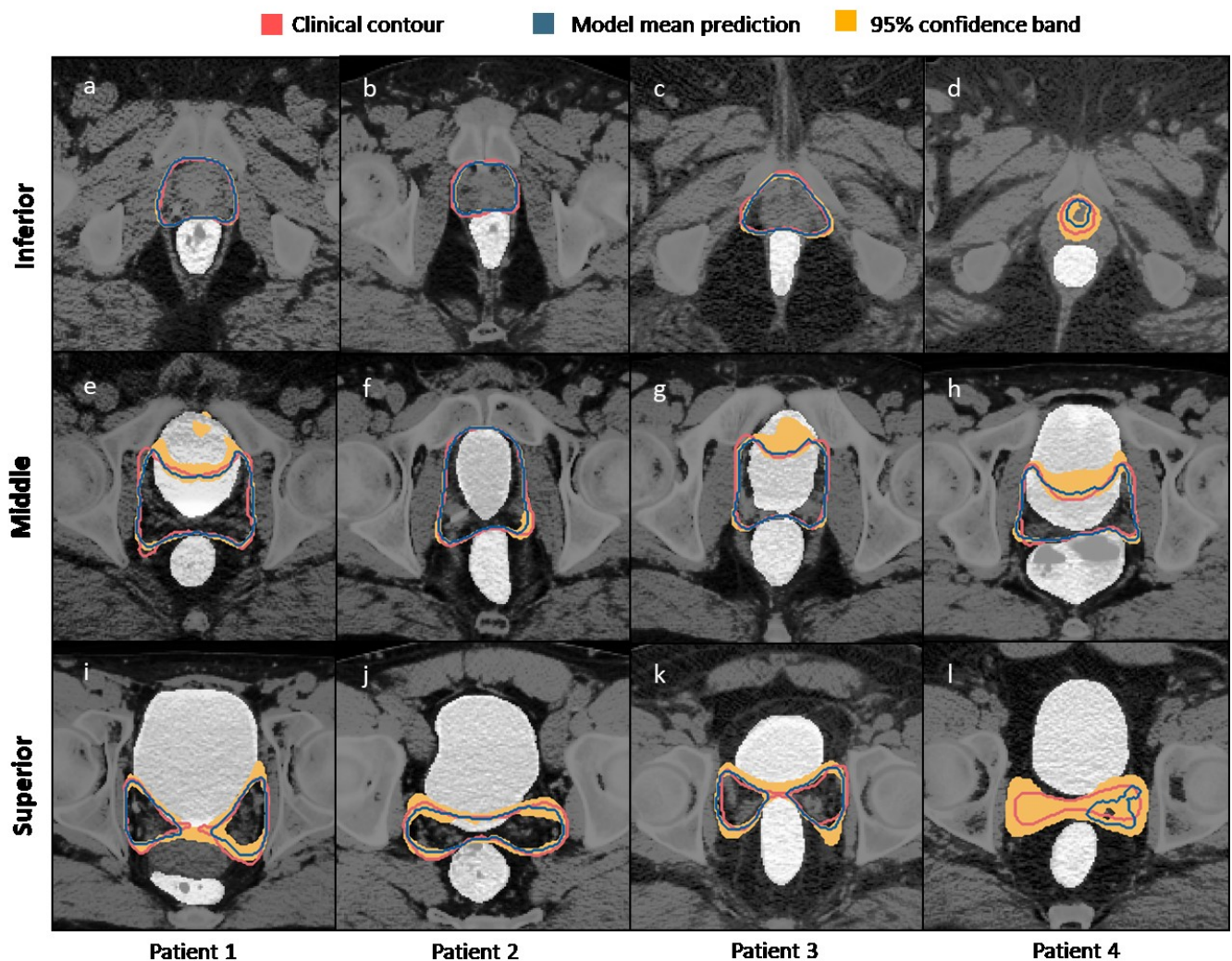

**Figure 3.**

*Visualization of the clinical CTV contours (red) and the predicted mean CTV contours (blue) with 95% confidence bounds (yellow) in axial CT images at three representative anatomical locations (top row - inferior, middle row – middle, and bottom row – superior) for four example testing patients (each column corresponding to one patient).*

We also used the results of the MCDO to calculate a metric to measure the overall quality of the segmented OARs. We defined *contour quality* as the mean DSC value between the each of the MCDO predictions and the mean model prediction for an OAR of each patient. We found that the contour quality correlates well ($R^2 = 0.89$) with the DSC value calculated between the model prediction and the clinical contour for OARs (**Figure 4**). Therefore, this quality metric can help physicians assess the model-predicted contours. We also applied this concept for CTVs but did not see a good correlation with DSC; this is probably attributable to the large inter-physician variations in CTV contouring.

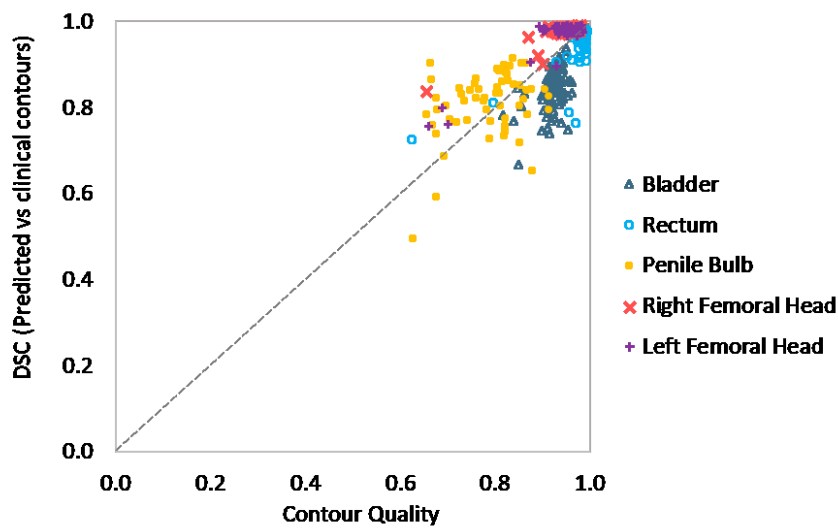

**Figure 4**.
*Scatter plot showing a correlation between the DSC values of the predicted OAR contours and the estimated Contour Quality ($R^2 = 0.89$).*

**Comparison with medical residents.** For each test patient, two out of five residents manually segmented the CTV with the assistance of pathology and MRI reports. We compared the DSC values between resident-drawn contours and clinical contours with the DSC values between the model contours and clinical contours for all test patients. We found that 87% of the time, the model outperformed the residents (**Figure 5a**). The average DSC for the model contours was statistically superior to the average DSC for the two residents by 7.1% (p-value <.0001) and to the better of the two residents by 5.1% (p-value <.0001) (**Figure 5b**). This indicates that our DL framework could be used either as a virtual resident to generate initial CTV contours to improve clinical workflow for attending physicians without residents, or as an education tool to guide residents in improving their CTV contours for attending physicians to review and revise.

**Evaluation of clinical acceptability.** Experienced practicing physicians were presented, in a randomized and blinded way, with the model-predicted CTV contour and the clinical CTV contour, side by side, for each of 30 anonymized patients selected from the 50 test patients who had treatments before 2019. These physicians, with the assistance of pathology reports, reviewed and scored the contours according to a 4-point grading system: 4 - acceptable without changes, 3 - acceptable with minor changes, 2 - acceptable with major changes, and 1 - completely unacceptable. Physicians were also asked to predict which of the two contours was the clinical contour used for patient treatment. Half of the patients were evaluated by their respective original treating physicians, and the rest were evaluated by a physician who was not involved in the original treatment.

For patients treated and scored by the same physician ("same-observer evaluation"), clinical contours were scored at an average of 3.4, and model-predicted contours were scored at 3.2. Physicians were able to correctly identify 60% of their own contours from AI contours. For patients treated and scored by different physicians ("different-observer evaluation"), clinical contours were scored at an average of 3.1, and model contours were scored at 3.3. Physicians were only able to correctly distinguish 25% of the clinical contours from the model contours. It appears that physicians were more confident in the clinical contours of their own patients, but less confident in those of other physicians' patients, compared to the contours produced by the model. However, separate statistical significance tests for single-observer and different-observer evaluations did not produce any statistically significant conclusions because of the small sample sizes. When mixing the single-observer and different-observer evaluations, the scores are 3.3 for both the clinical and model-predicted contours, and an equivalence test for means showed that they are equivalent with p-values < 0.05 for both. None of the contours received a score below 2. The scores are summarized in **Figure 5c**.

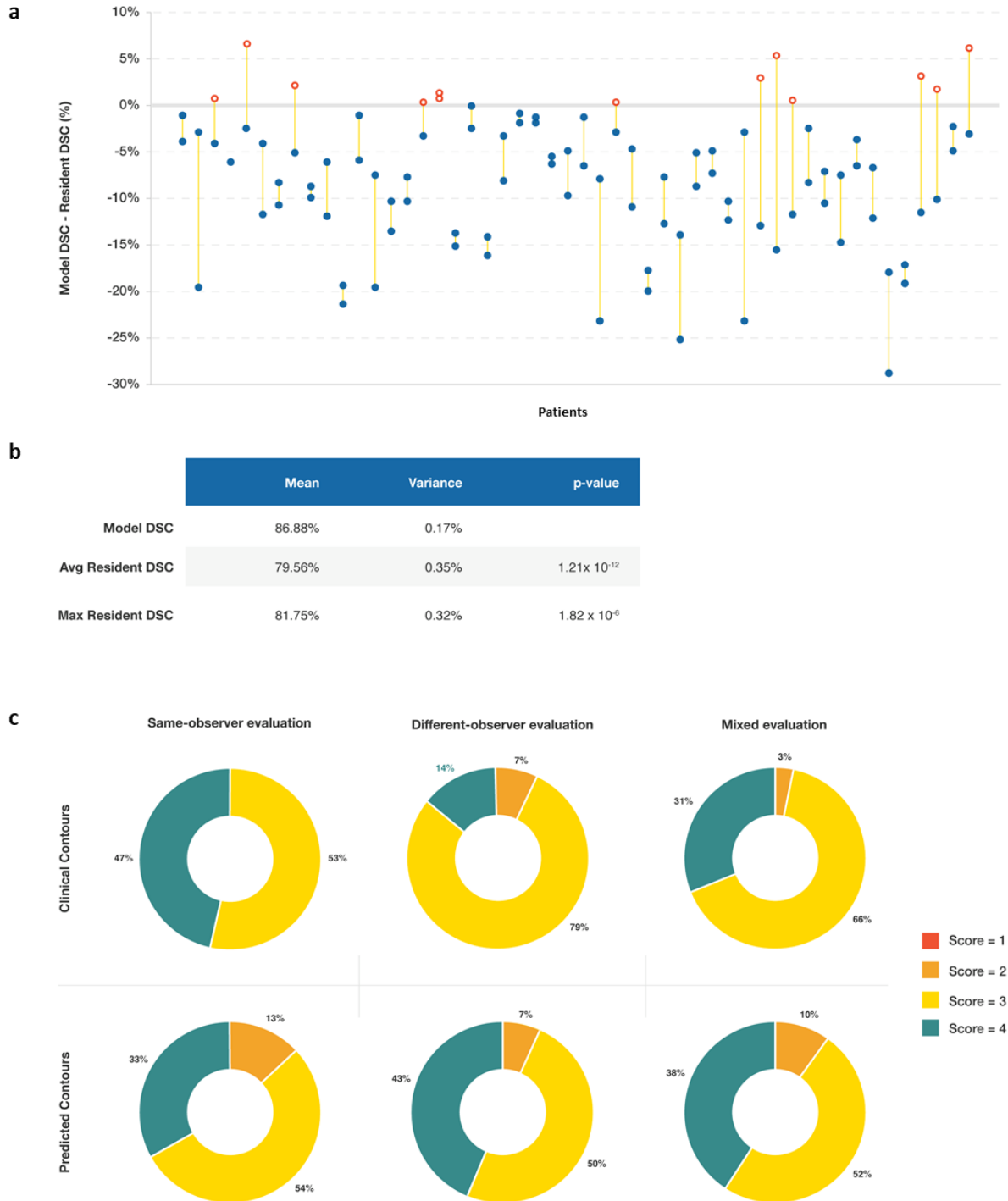

*Figure 5.*

*Comparisons between manual and model-predicted CTV contours. **a**, Comparison between resident contours and model-predicted CTV contours using the clinical contours as references for 50 test patients. For each patient, the two circles represent the results from the two residents. Red open circles mean the residents outperformed the model; blue filled circles mean the model outperformed the residents. **b**, The mean and variance of DSC values of the model-predicted CTV contours, the average of the two residents, and the better of the two residents for each patient. p-values for the t-test are also shown. **c,** Summary of the* *evaluation of clinical acceptability using a 4-point grading system. None of the contours received a score below 2.*

## Discussion

Our first contribution in this work is that we developed a new DL-based framework for fully automated segmentation of CTVs and OARs on post-operative prostate cancer CT images. We demonstrated that the framework is effective for post-operative CTV segmentation, as it learns expert domain knowledge from clinical contours to extract the most appropriate discriminative features even in the absence of a visible segmentation volume. The proposed OAR segmentation models perform with high accuracy for all the structures. The CTV segmentation model can generate a DSC of 87%, which is similar to the performance of state-of-the-art CTV prediction models that work with visible GTVs [21,22,23]. The success of our model can be attributed to its three-stage design, which localizes structures, segments OARs, and then uses the segmented OARs along with the localized CTV VOI to automatically segment the CTV. The structure localization step helps by localizing VOIs so that image features can be extracted around these focal areas. This step also enables the use of adaptive histogram normalization within each VOI, which we found affects the segmentation accuracy. The initial localization step also cuts down GPU memory consumption, as fitting an entire 3D CT volume is computationally intensive. By choosing separate architectures to segment each OAR in the second step, we demonstrated that all of the OARs could be segmented with high accuracy. Since post-operative CTV segmentation relies on the surrounding anatomical borders, using segmented OARs along with the CTV VOI as model inputs proved effective. In addition, because the CTV segmentation depends on anatomical location, adding distance prediction as an auxiliary task helped the model to extract features more effectively.

Our second contribution is that our model estimates the segmentation uncertainties, which facilitates clinical implementation of the developed DL framework. The uncertainty associated with the segmentation networks can be measured by exploiting a relationship between MCDO and a Bayesian posterior. Even though all the physicians and residents follow the same institutional guidelines for segmenting post-operative prostate CTVs, there is significant variability among these human experts because there is no visual ground truth. Clinical guidelines give the floor but not the ceiling, so there is still room for physicians to exercise their own clinical judgements based on their experience, knowledge, and training background, which leads to high variability in the final CTV contours. This kind of variability is unlikely to diminish in the near future because its effect does not easily manifest among the many confounding factors in the clinical outcome and toxicities. This being the case, one must consider the large uncertainties in particular areas of the CTV contours when implementing a DL-based automatic segmentation framework in clinical practice. Our study found (**Figure 5c**) that, for most patients, the CTV contours require minor revisions before they can be used for patient treatment. It would be tedious and time consuming for physicians to go through every CT slice and inspect every part of contours, which would make our DL tool unacceptable for clinical application. We proposed presenting physicians the 95% confidence bounds together with the predicted mean contours. We found that when the estimated uncertainty is small, the contours predicted by the model match well with the clinical contours. In the areas where the uncertainty is large, the predicted contours may or may not agree well with the clinical contours. Therefore, physicians can focus on the areas of the contours with wide 95% confidence bounds. This will probably improve efficiency and perhaps even accuracy when physicians use the DL tool to assist in segmenting CTVs for post-operative prostate cancer radiotherapy.

Our third contribution is that we demonstrated the clinical acceptability of the developed DL framework by comparing it with medical residents and with clinical contours through a reader study. In current clinical practice, residents typically draw the CTV contour first, then attending physicians review and revise them. Our clinical evaluation study shows that the DL framework we developed can outperform residents, which suggests that it could be used to guide residents in generating the initial CTV contours for physicians to review and revise. Alternatively, for attending physicians without residents, the DL framework could be used as a virtual resident to generate initial CTV contours to improve clinical workflow. Although the reader study shows no significant difference between the clinical contours and the DL contours with respect to quality, it still suggested using the DL tools to assist, not to replace, physicians for segmenting CTV contours for post-operative prostate cancer radiotherapy.

## Methods

**Localization of OARs and CTVs.** We use a 2D U-Net [31] architecture with 5 output channels (corresponding to CTV, bladder, rectum, femoral heads, and penile bulb) to localize CTVs and OARs before 3D segmentation. This step is necessary because CT volumes are 512×512×~200 in size, so it is difficult to fit the entire volume into a 3D network without downsizing, which would lower the resolution. The 2D localization network gives coarse segmentations of the CTV and OARs.

We use two loss functions, $L_1$ and $L_2$, to train the localization network. Apart from playing an important role in resolving class imbalance problems, the loss functions used to train the localization network directly control the trade-off between sensitivity and specificity. $L_1$ is the negative of DSC, the most commonly used loss function in segmentation:

$$L_1 = -\frac{2\sum_{i=1}^{N} w_i p_i q_i}{\sum_{i=1}^{N} w_i p_i + \sum_{i=1}^{N} w_i q_i} \tag{1}$$

Here, $p_i \in [0,1]$ and $q_i = 0,1$ are the predicted and ground truth values, respectively, of the *i*-th pixel, *N* is the total number of pixels in the image, and $w_i$ is the weighting factor for the *i*-th pixel. $w_i$ is set to 1 for the localization network and thus is not shown in Equations (2)-(4). The partial derivative of $L_1$ during gradient backpropagation is

$$\frac{\partial L_1}{\partial p_j} = -2\left[\frac{q_j\left(\sum_{i=1}^{N} p_i + \sum_{i=1}^{N} q_i\right) - \sum_{i=1}^{N} p_i q_i}{\left(\sum_{i=1}^{N} p_i + \sum_{i=1}^{N} q_i\right)^2}\right] \tag{2}$$

With this loss function, foreground pixels ($q_j = 1$) get larger absolute gradient values and thus, are weighted higher than the background pixels ($q_j = 0$), which ensures that the network performs well even in cases of class imbalance. However, this loss function does not stress incorrectly classified foreground pixels, which directly relates to sensitivity. To increase the sensitivity, we designed another loss function $L_2$:

$$L_2 = -\frac{2\sum_{i=1}^{N}\sqrt{(p_i + \varepsilon)}\, q_i}{\sum_{i=1}^{N}\sqrt{(p_i + \varepsilon)} + \sum_{i=1}^{N} q_i}, \tag{3}$$

where $\varepsilon$ is a small positive number. The partial derivative is

$$\frac{\partial L_2}{\partial p_j} = -(p_j + \varepsilon)^{-1/2} \left[ \frac{q_j(\sum_{i=1}^{N}\sqrt{(p_i + \varepsilon)} + \sum_{i=1}^{N} q_i) - \sum_{i=1}^{N}\sqrt{(p_i + \varepsilon)}q_i}{(\sum_{i=1}^{N}\sqrt{(p_i + \varepsilon)} + \sum_{i=1}^{N} q_i)^2} \right] \quad (4)$$

The additional term $(p_j + \varepsilon)^{-1/2}$ in the gradient ensures that large gradients are assigned to foreground voxels that are misclassified. But this also reduces specificity because less attention is paid to the hard samples in the background. The localization network is first trained with $L_1$, then fine-tuned in the last 10 epochs with $L_2$ to reduce the number of false positives.

Even though class imbalance is managed using the loss functions, there is still a large dataset imbalance due to the large number of unlabeled 2D CT slices. To ensure effective training, we applied affine transformations to the labeled slices and randomly abandoned unlabeled slices to balance the dataset and improve the learning process. We used the centroids of the predicted coarse contours to crop the original CT images into VOIs for the CTV and OARs.

**Segmentation Networks.** The VOIs for bladder, rectum, right femoral head, left femoral head and penile bulb are used as inputs for the organ segmentation networks. Because of the differences in OARs' shapes and the intensity variations within them, we use different DL architectures to individually segment each OAR. We use a 3D UNet architecture with ResNeXt [32] blocks in the encoding arm for bladder and for femoral heads, a 3D UNet architecture with ResNet [33] and inception blocks [34] in the encoding arm for rectum, and a 3D UNet with squeeze and excitation blocks [35] for penile bulb segmentation.

Since the CTV is defined by the anatomic boundaries of nearby OARs, the inputs of the DL network for CTV segmentation are the CTV VOI and the segmented bladder and rectum masks filled with the corresponding CT numbers. The encoding path of the U-Net is connected to two parallel but structurally similar decoding paths, one for segmentation map prediction and the other for distance map prediction (**Figure 6**). We were motivated to introduce the distance map prediction by the success of self-supervised methods that learn deep features by creating multiple auxiliary tasks [36]. The distance mask for training is created using the Euclidean distance transform of the binary mask, by assigning zero value to the pixels inside the CTV and the minimum distance from that pixel to the CTV boundary to the pixels outside the CTV. This introduces the additional task of learning the anatomical location. Since we are using a 3D deep neural network, to efficiently backpropagate the loss function to shallow layers, we enable deep supervision by injecting two additional segmentation predictors to the network's segmentation path. The loss function used to train the CTV segmentation network has four components: 1) $L_1$ as defined in Equation (1) for the main segmentation prediction, with $w_i = 0.8$ for boundary and 0.2 for non-boundary voxels; 2) and 3) $L_1$ for each of the two auxiliary segmentation predictions, with $w_i = 1$ for all voxels; and 4) a mean squared error loss for the distance map prediction.

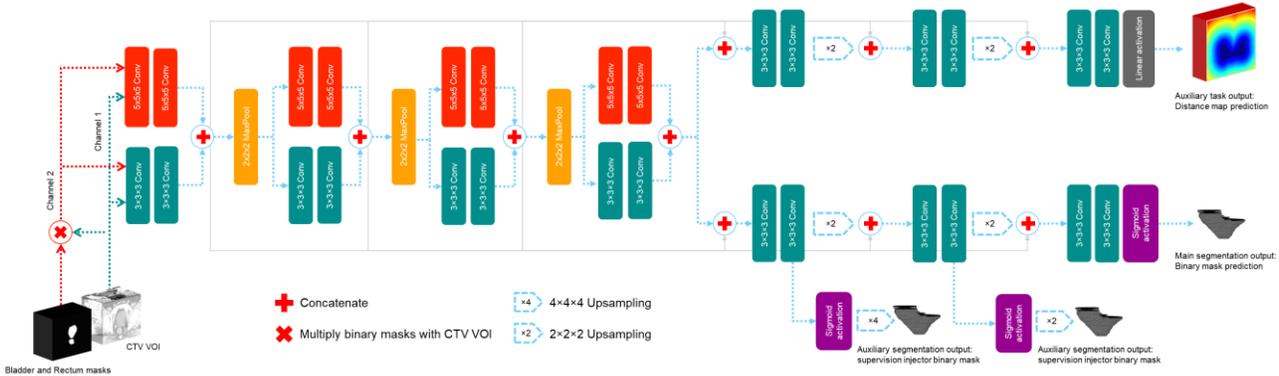

*Figure 6:*
*Anatomy-guided multi-task network (AG-MTN) for CTV segmentation taking the CTV VOI and segmented bladder and rectum masks as inputs. The outputs of the network are a main segmentation prediction, two auxiliary segmentation predictions, and an auxiliary distance map prediction.*

For the 3D OAR and CTV segmentation networks, we randomly applied data augmentation techniques such as rotating by small angles (<10°), image scaling, and image flipping during training for more effective learning. We used adaptive histogram equalization for data preprocessing to enhance edge definitions and improve local contrast.

**Uncertainty and quality estimation.** By using MCDO, where subsets of a network are inactivated during training to avoid overfitting, one can compute an approximation of the posterior distribution by sampling multiple predictions with dropout turned on. Dropout at test time approximates sampling from a Bernoulli distribution over the network weights. This allows one to perform approximate but efficient Bayesian inference by implementing existing software in a straightforward way. Because of fast inference with CNNs, multiple MCDO samples can be generated to reliably approximate the posterior distribution in an acceptable time.

Although dropout [38] is widely used as a regularization technique for fully connected layers, it is often less effective for convolutional layers where features are correlated spatially. For estimating uncertainty in this segmentation work, we use Monte Carlo sampling with DropBlock [39], a structured form of dropout, to better regularize convolutional networks. By leaving DropBlock turned on at the test time, we can draw multiple Monte Carlo samples from the approximate predictive posterior. Since MCDO predictions approximate a Gaussian distribution, along with a mean prediction, they can also be used to calculate an upper and lower 95% bound.

### Data availability

All the datasets were collected from one institution and are non-public. In accordance with HIPAA policy, access to the datasets will be granted on a case by case basis upon submission of a request to the corresponding authors and the institution.

### Code availability

The DL models are free to download for non-commercial research purposes on GitHub. (https://github.com/anjali91-DL/Post-op-prostate-DL-model)

**Author Contributions**

AB and SJ conceived the original idea. AB, DN and SJ designed the project and experiments. AB and DN developed and implemented the DL networks and performed the statistical analysis. AB conducted validation experiments. AB, HM, MD, YW and ML collected and curated patient data. AG, ND, RH, HM, MD, YW and ML helped with the clinical aspects and performed the clinical analysis. YG and ASB contributed to the software development. AB, DN and SJ prepared the manuscript. HM, AG and MD reviewed and edited the manuscript. SJ provided supervision and funding for the project. All authors discussed the results and reviewed the manuscript.

**Acknowledgement**

We would like to thank Dr. Jonathan Feinberg for editing the manuscript and Varian Medical Systems, Inc. for providing funding support.